\documentclass[a4paper,12pt]{article}
\usepackage{amsmath, amsthm, amsfonts, amssymb}
\usepackage[svgnames]{xcolor}
\colorlet{color1}{NavyBlue}
\usepackage[colorlinks=true,allcolors=color1]{hyperref}
\usepackage{physics}
\usepackage{dsfont}
\usepackage{graphicx}
\usepackage{cleveref} %use \cref{label}
\usepackage{xfrac}
\usepackage{cite}

\usepackage[margin=1in]{geometry}

\title{\vspace*{-1cm} \par \Large\bf Stellar equilibrium on a physical vacuum soil}
\author{
Julio Arrechea$^{\dag}$ and 
Carlos Barcel\'{o}$^{\ddag}$
\vspace{0.4cm}
\\ 
\small\it  Instituto de Astrof\'{i}sica de Andaluc\'{i}a (IAA-CSIC),
 Glorieta de la Astronom\'ia, \\ \small\it18008 Granada, Spain}
\date{}

\begin{document}
%--------------------------------------------------
\maketitle
%--------------------------------------------------
%\thispagestyle{empty}
%--------------------------------------------------
%-------------------------------------------------------------------------
\begin{center}
Dated:\date{{ \today}}
\end{center}
%------------------------------------------------------------------------------
\vspace{0.4cm}
%-----------------------------------------------------------------------
\hrule
\begin{abstract}
\noindent
We show that the repulsive effects associated to the zero-point energies of quantum fields are capable of supporting ultracompact stars that overcome the compactness limits present in general relativity for any object in hydrostatic equilibrium. These objects are exact self-consistent solutions in semiclassical gravity that incorporate the backreaction of the renormalized stress-energy tensor (RSET) of quantum fields in vacuum. We arrive at stars of striking qualitative agreement through two independent modelings of the RSET, evidencing the generality and robustness of this result. The main physical properties of these novel black hole mimickers are reviewed.
\end{abstract}
%-----------------------------------------------------------------------
\vspace{0.1cm}
\hrule
\vspace{0.3cm}
%\maketitle
\begin{center}
{\centerline{Essay written for the Gravity Research Foundation 2023 Awards for Essays on Gravitation}}
\end{center}
%---------------------------------------------------------------------

%-----------------------------------------------------------------------
{$^\dag$\underline{Electronic address}: arrechea@iaa.es}

{$^\ddag$\underline{Electronic address}: carlos@iaa.es
} \\
{Corresponding author: Julio Arrechea}
%-----------------------------------------------------------------------

\thispagestyle{empty}
\clearpage
%----------------------------------------------------------------------
\markright{ \hfil }
%------------------------------------------------------------------------------
\pagestyle{myheadings}
%------------------------------------------------------------------------------

%----------------------------------------------------------------
\def\HRULE{{\bigskip\hrule\bigskip}}
%----------------------------------------------------------------

\setcounter{page}{1}
%---------------------------------------------
\section*{Introduction}
%---------------------------------------------
It is by now absolutely clear that the cosmos is spattered by very dark and compact objects.
Probes of the shape of spacetime around these objects show that they are compatible with the Kerr metric (see for example~\cite{EventHorizonTelescope2022}). However, at most, current observations have not penetrated further 
than $10^{-4}$ times the gravitational radia~\cite{Carballoetal2018c,Carballo-Rubioetal2022} (going up to $10^{-15}$ in the case of Sagitarius A$^*$) and typically they are not free from theoretical hypotheses~\cite{Carballoetal2018c}.
Given the situation, it is fair to say that precise tests of the Kerr hypothesis are yet to be achieved. Notwithstanding this, the extended belief that these objects are {\it necessarily} general relativity black holes is not only coming from the observations, but strongly conditioned by an additional theoretical argument: In general relativity there are no stable stellar configurations surpassing the Volkov-Oppenheimer limit for nuclear matter~\cite{OppenheimerVolkoff1939} nor the Buchdahl limit limit for any sensible matter source~\cite{Buchdahl1959}.

In this essay we show that this theoretical argument is not as strong as it would seem. It weakens just by taking into account that the standard vacuum of general relativity is {\it too empty} to represent the physical quantum vacuum.
We present the most robust evidence to date that only incorporating vacuum polarization effects as an additional source in the Einstein equations permits the existence of horizonless ultracompact stars. By ultracompact we mean objects with compactness extremely close to black holes, hence amply surpassing the Buchdahl limit. 
For this, we solve the semiclassical Einstein equations
%------------------
\begin{equation}\label{Eq:SemiEinstein}
    G^{\mu}_{\nu}=8\pi G\left(T^{\mu}_{\nu}+\langle\hat{T}^{\mu}_{\nu}\rangle\right),
\end{equation}
%------------------
sourced by a constant-density classical fluid (we expect our result not to depend crucially on this idealized content), and incorporating the renormalized stress-energy tensor (RSET) of a single massless, minimally coupled scalar field. Remarkable effort is already required to calculate the RSET of scalar fields in the Schwarzschild~\cite{HowardCandelas1984}, Reissner-Nordström~\cite{Andersonetal1994}, and Kerr black holes~\cite{Levietal2016}, not to mention solving their corresponding backreaction problems. Even when the RSET is obtainable in closed form, it is acknowledged that the semiclassical equations are a system of high-derivative-order differential equations which carry along further pathologies~\cite{Simon1990,FlanaganWald1996}. This is why, in order to explore the physics concealed in~\eqref{Eq:SemiEinstein} in a systematic way, it is customary to develop approximate RSETs (adapted to spherically symmetric stars in this case) that preserve qualitatively the physics of exact RSETs while being free from their shortcomings.

To derive semiclassical equations of second differential order, we follow two approximation schemes for the RSET, namely, the Regularized Polyakov approximation~\cite{Polyakov1981,Arrecheaetal2021} and the perturbative order reduction of the Anderson-Hiscock-Samuel RSET~\cite{Andersonetal1994,Thesis}, that are entirely based on distinct physical grounds. Most notably, these two unrelated RSETs give rise to ultracompact stars of striking qualitative (and, to a large extent, quantitative) similitude, manifesting the robustness of semiclassical analyses and providing independent proofs of the existence of these objects. We turn now towards outlining the derivation of semiclassical ultracompact stars in these two frameworks, omitting here the characteristics of the external geometry of these stars, for which we refer the reader to~\cite{Arrecheaetal2019,Arrecheaetal2022b}.

%---------------------------------------------
\section*{Semiclassical hydrostatic equilibrium}
%---------------------------------------------
We consider static and spherically symmetric spacetimes described by the line element
\begin{equation}\label{Eq:4DMetric}
    ds^{2}=-f(r)dt^{2}+h(r)dr^{2}+r^{2}d\Omega^{2}
\end{equation}
where $d\Omega^{2}$ is the line element of the unit sphere and we define \mbox{$C(r)\equiv2m(r)/r=1-h(r)^{-1}$} as the compactness function, with $m(r)$ the Misner-Sharp mass~\cite{MisnerSharp1964,HernandezMisner1966}. 
We model the stellar interior via the stress-energy tensor (SET) of an isotropic perfect fluid
%----------------------------------------------------------
\begin{equation}\label{Eq:ClassicalSET}
    T^{\mu}_{\nu}=\left(\rho+p\right)u^{\mu}u_{\nu}+p\delta^{\mu}_{\nu},
\end{equation}
%----------------------------------------------------------
with $p$ and $\rho$ denoting the pressure and energy density measured by an observer comoving with the fluid with $4$-velocity $u^{\mu}$. Covariant conservation of the SET~\eqref{Eq:ClassicalSET} imposes the relation
%----------------------------------------------------------
\begin{equation}\label{Eq:Cont}
    \nabla_{\nu}T^{\mu}_{r}=p'+\frac{f'}{2f}\left(\rho+p\right)=0.
\end{equation}
%----------------------------------------------------------
Additionally, we need to impose an equation of state for the classical fluid. We assume a simple uniform energy density fluid
\begin{equation}\label{Eq:EoS}
    \rho(r)\equiv\rho=\text{const},
\end{equation}
so that we can find the semiclassical counterparts of Schwarzschild's stellar solutions~\cite{Schwarzschild1916}.

The parameter \mbox{$C(R)\in(0,1)$} denotes the compactness of a given star at its surface \mbox{$r=R$}. In general relativity, the Buchdahl theorem imposes an upper bound of $C(R)=8/9$ to the compactness of any object in hydrostatic equilibrium~\cite{Buchdahl1959}. This bound contributes to the established consensus that observed massive bodies whose compactness nears the black hole limit $C(R)=1$ are presumably black holes. Among other hypotheses~\cite{UrbanoVeermae2018}, the Buchdahl theorem assumes the energy density of a star to be an outwards-decreasing function of the radial coordinate $r$, a condition saturated
by the equation of state~\eqref{Eq:EoS}. In the spacetimes~\eqref{Eq:4DMetric}, however, the RSET corresponds to an anisotropic perfect fluid that can contribute negatively to the total energy density of highly compact stars~\cite{Hiscock1988,ReyesTomaselli2023}. When the RSET (in any of the approximate forms presented below) is allowed to backreact on the star, these negative energies allow to surpass the Buchdahl limit. Semiclassical gravity thus becomes an outstanding theory in which to search for horizonless alternatives to black holes.

%----------------------------------
\subsection*{{\rm{A}}. Stellar equilibrium in the Polyakov approximation}
%----------------------------------

Essential features of the propagation of a massless minimally coupled scalar in four spacetime dimensions can be captured by two-dimensional models, described by the $\{t,r\}$ sector of the line element~(\ref{Eq:4DMetric}).
The simplicity that stress-tensor regularization acquires in this dimensionally-reduced spacetime allows to express the corresponding RSET in closed analytic form~\cite{DaviesFulling1977}, which is then identified with a four-dimensional RSET through 
\begin{equation}\label{Eq:2Dto4D}
\langle\hat{T}^{\mu}_{\nu}\rangle^{\rm P}=F(r)\delta^{\mu}_{a}\delta^{b}_{\nu}\langle\hat{T}^{a}_{b}\rangle^{\text{(2D)}}
+(T_{\rm AC})^{\mu}_{\nu}
\end{equation}
where Greek and Latin indices take $4$ and $2$ values,
and $\rm AC$ stands for angular conservation. Here, $F$ is an arbitrary radial function to be specified by hand and which condenses some physical properties absent in the RSET upon uplifting it to four dimensions (mainly, its behavior at $r=0$). Lastly, the term $(T_{\rm AC})^{\mu}_{\nu}$ contains angular pressures that follow from requiring $\langle\hat{T}^{\mu}_{\nu}\rangle^{\rm P}$ to be conserved,
\begin{align}\label{Eq:ConsRSET}
    \nabla_{\mu}\langle\hat{T}^{\mu}_{r}\rangle^{\rm P} =
    ~\partial_{r}\langle\hat{T}^{r}_{r}\rangle^{\rm P} +\frac{2}{r}\left(\langle\hat{T}^{r}_{r}\rangle^{\rm P} -\langle\hat{T}^{\theta}_{\theta}\rangle^{\rm P} \right)
    +\frac{f'}{2f}\left(\langle\hat{T}^{r}_{r}\rangle^{\rm P} -\langle\hat{T}^{t}_{t}\rangle^{\rm P} \right)=0.
\end{align}
This way we arrive at the Regularized Polyakov RSET,
\begin{align}\label{Eq:PolyakovRSET}
    \langle\hat{T}^{t}_{t}\rangle^{\rm P}=
    &
    \frac{F}{96\pi h}\left[\frac{2f'h'}{fh}+3\left(\frac{f'}{f}\right)^{2}-\frac{4f''}{f}\right],\nonumber\\
    \langle\hat{T}^{r}_{r}\rangle^{\rm P}=
    &
    -\frac{F}{96\pi h}\left(\frac{f'}{f}\right)^{2},\nonumber\\
    \langle\hat{T}^{\theta}_{\theta}\rangle^{\rm P}=
    &
    -\frac{\left(2F+rF'\right)}{192\pi h}\left(\frac{f'}{f}\right)^{2},
\end{align}
 which is unique up to specifying $F$. The traditional choice is $F(r)=1/ r^{2}$, which eliminates angular contributions and perfectly captures the physics around black hole horizons~\cite{ParentaniPiran1994,Fabbrietal2005,Chakrabortyetal2015}. Applied to stellar situations, however, this choice of $F$ yields a singular RSET at $r=0$ and needs to be refined further, ideally, so that the Regularized Polyakov RSET most closely resembles the exact RSET near $r=0$. From here onwards, we will use the traditional $F$ except inside a regularizing core.

\begin{figure} 
\includegraphics[width=\textwidth]{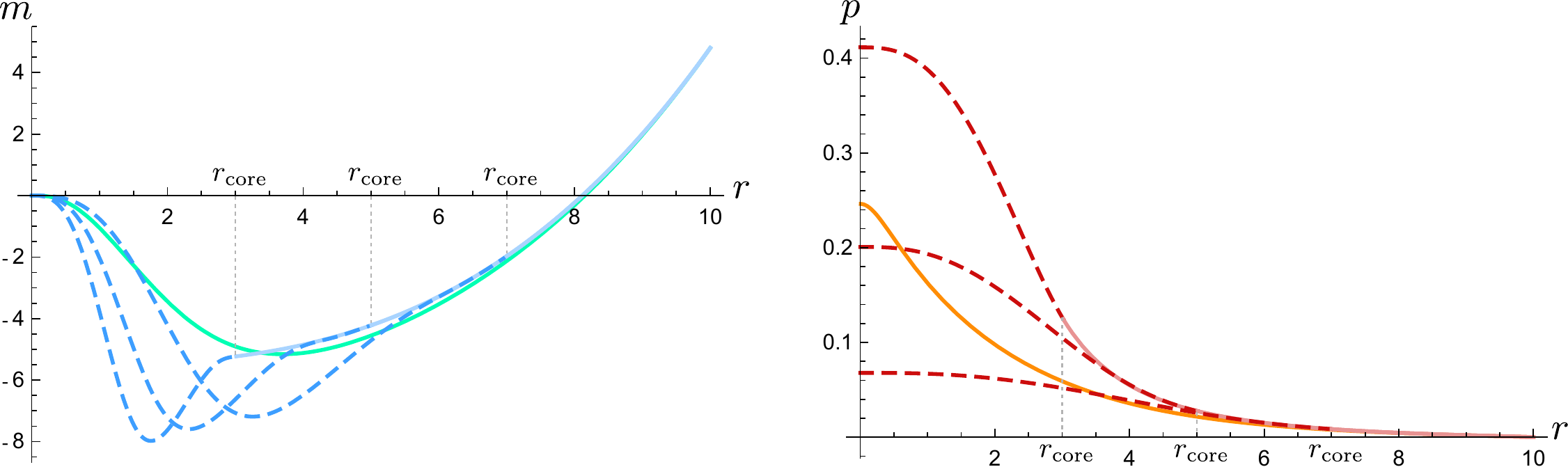}
    \caption{Misner-Sharp mass (left plot) and classical pressure (right plot) of stars with $C(R)=0.96$ and $R=10$. The dashed curves represent solutions obtained through the Regularized Polyakov approximation. We have shown three solutions with $\rho\simeq0.00246$ and different $r_{\text{core}}$ values, below which the $F$ function is distorted. We have also plotted a star obtained in the order-reduced prescription with $\rho\simeq0.00248$ (turquoise and orange lines) that highlight the qualitative agreement obtained through both RSET approximations.}
    \label{Fig:Stellar}
\end{figure}

To arrive at this result, we followed a reverse-engineering process (details available in~\cite{Arrecheaetal2022}), first matching the classical pressure at $r_{\text{core}}$ with a regular, analytic pressure profile (the simplest assumption among all possibilities), to then solve the semiclassical equations for $F$. We find broad families of $F$ functions that result in regular compact stars that surpass the Buchdahl limit. These stars exist for any choice of $r_{\text{core}}$ and their Misner-Sharp mass and classical pressure are displayed in Fig.~\ref{Fig:Stellar}.

The viability of the standard Polyakov approximation, though very good in many situations, is limited in stellar spacetimes due to the ambiguities inherent to dimensional reduction. 
However, we should not detract the fact that, within the space of possibilities allowed by our regularization procedure, it is possible to find that the effects of quantum vacuum polarization can support stars surpassing the Buchdahl limit. 
To erase all charges of {\it ad hoc} procedure,
in the next section we show that we arrive at similar conclusions through an entirely unrelated RSET approximation which is four dimensional from the start, thus naturally adapted to stellar spacetimes.

%--------------------------------
\subsection*{ {\rm B.} Stellar equilibrium in the order-reduced prescription}
%--------------------------------
The RSET of a scalar field in four dimensions was derived by Anderson, Hiscock and Samuel~\cite{Andersonetal1994}.
They found it admits a decomposition into two independently conserved, analytic and numeric parts. The analytic part, named AHS-RSET hereafter, can be used as an approximation
to the full RSET of massless fields, with the drawback that it contains high-order spatial derivatives of the metric that entail the appearance of spurious solutions much like in the Abraham-Lorentz equation~\cite{LandauLifschits1975}. Inspired by the treatment followed with the radiation reaction problem and semiclassical corrections in cosmology~\cite{ParkerSimon1993,FlanaganWald1996}, we apply a perturbative reduction of order scheme to the semiclassical equations to cast them into a system where high-derivative-order terms appear at $\order{\hbar^{2}}$, eliminating an important fraction of these spurious solutions.

The procedure of perturbative reduction of order is as follows. Take the $tt$ and $rr$ components of the semiclassical equations and neglect terms of $\order{\hbar}$, such that
\begin{align}\label{Eq:beyond:hbarExpMat}
    \frac{h(1-h)-rh'}{h^{2}r^{2}}=
    &
    -8\pi\rho+~\order{\hbar},\nonumber\\
    \frac{rf'+f-fh}{fhr^{2}}=
    &
    8\pi p+~\order{\hbar}.
\end{align}
Consecutively differentiating these expressions and replacing derivatives of $p$ by derivatives of $f$ through Eq.~\eqref{Eq:Cont}, we derive a set of expressions relating derivatives of any order of $f,h$, and $p$, with these functions themselves.
Upon replacing said relations in the AHS-RSET, its $\langle\hat{T}^{t}_{t}\rangle$ and $\langle\hat{T}^{r}_{r}\rangle$ components are reduced to expressions without derivatives. Lastly, we obtain low-derivative-order expressions for the $\langle\hat{T}^{\theta}_{\theta}\rangle$ component by imposing covariant conservation through~\eqref{Eq:ConsRSET}, in a manner analogous to the Regularized Polyakov RSET. This leads to the Matter-Order-Reduced RSET (MOR-RSET), a semiclassical source that is well defined at $r=0$. Full expressions are available in~\cite{Thesis}. We avoid showing them here because they are lengthy and opaque.

We assume a minimally coupled scalar field to allow for a faithful comparison with the Regularized Polyakov approximation and leave non-minimally-coupled fields for future studies. Sourced by the MOR-RSET,
the semiclassical equations are of the same derivative order as the classical Einstein equations and can be solved as a boundary value problem. By specifying the integration parameters $R$,~$C(R)$ and $\rho$, we perform an exploration of the space of solutions and find regular stars that surpass the Buchdahl limit. The characteristics of these solutions, in comparison with those found with the Regularized Polyakov approximation (see Fig.~\ref{Fig:Stellar}), are described in the next section.

%---------------------------------------------
\section*{Physical properties of semiclassical black stars}
%---------------------------------------------
We have followed two unrelated approaches to modeling the RSET of a massless, minimally coupled scalar field, whose backreaction allows, in both cases, for the existence of a new type of ultracompact object supported by quantum vacuum polarization. The robustness of this result is manifested through its generality, as both approaches not only agree in the existence of semiclassical black stars, but also on their most basic properties. What makes their matter overcome gravitational contraction is the presence of a negative mass interior, generated by the negative energy densities characteristic of the vacuum. These negative masses exert an additional gravitational repulsion that supports the object, while
relation~\eqref{Eq:Cont} translates their inwards-increasing classical pressures into large interior redshifts. The reader can see, in Fig.~\ref{Fig:Stellar}, some exact solutions obtained through both approximations and that surpass the Buchdahl limit.

\begin{figure}
    \centering
    \includegraphics[width=0.65\textwidth]{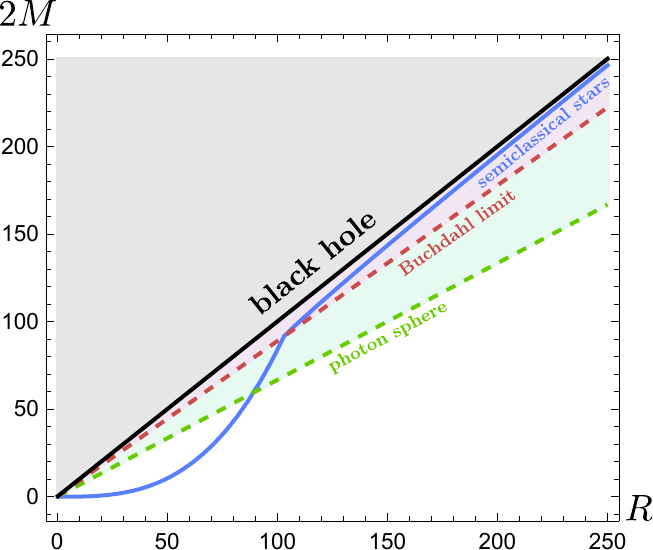}
    \caption{Mass-to-Radius diagram of semiclassical stars in the Regularized Polyakov approximation with $\rho=10^{-5}$ (in Planck units). The black line represents the compactness parameter of black holes, $C(R)=1$, the dashed red line denotes the Buchdahl compactness bound \mbox{$C(R)=8/9$}, and the dashed green line is the minimum compactness of objects that exhibit a photon sphere, $C(R)=1/3$. The blue curve represents semiclassical stars. For those stars surpassing the Buchdahl limit, the total mass $M$ grows approximately linearly with the radius $R$. Each point within the blue curve admits whole families of regulator functions $F$ for which the entire geometry is regular. A qualitatively similar diagram is obtained making use of the MOR-RSET.}
	\label{Fig:MtoRDiagram}
\end{figure}

Semiclassical stars span a wide range of compactness values. Their Mass-to-Radius diagram (Fig.~\ref{Fig:MtoRDiagram}) shows three distinct regimes of solutions. For $C(R)<8/9$, these stars are perturbatively corrected constant-density stars nearly undistinguishable from their classical counterparts. Around the Buchdahl limit a drastic transition occurs as negative energies start building up in the central regions of the object. For \mbox{$1> C(R)>8/9$} we have regular stars sustained by a negative mass core.

Semiclassical black stars exhibit a photon sphere, so they could produce gravitational wave echoes. To give a rough estimate, for the star obtained through the MOR-RSET model from Fig~\ref{Fig:Stellar}, the crossing time for a null ray emitted from the photon sphere to be reflected at $r=0$ and return to the photon sphere is
one order of magnitude larger (in units of the stellar mass $M$) than the time it takes to go from the photon sphere to the surface, i.e.,
\begin{equation}\label{Eq:CrossingTime}
    \tau_{\text{echo}}= 2 \int_{0}^{r_{\text{ph}}}\left(h/f\right)^{1/2}dr' \simeq 306 M \sim 52 \tau_{\text{s}},
\end{equation}
where $\tau_{\text{s}}$ is the light-crossing time between the photon sphere and the star surface.
Attending only to the geometrical features, this effect could result in gravitational-wave echoes clearly distinguishable from the characteristic signals associated to black hole mergers. 

It is clear that the RSETs we have used in our investigations correspond to treatable qualitative models of any average vacuum energy present in Nature (this should incorporate at least all the standard model of particles with all its interactions). Nonetheless, they illustrate how only considering a more realistic description of empty space, in tune to what we know about the quantum world, makes viable the existence of ultracompact stars which could be mistaken by black holes.
Even without knowing the precise mechanism by which these object might form, their very existence makes the {\it necessarily} argument for black holes a lot weaker. 
Our results reinforce the already clear need to perform precise test of the actual geometry of astrophysical black holes.
Future large-baseline-interferometry observations~\cite{EventHorizonTelescope2019} and gravitational wave observations~\cite{LIGOScientific2016}, in particular through the search of gravitational-wave echoes~\cite{Abedietal2016},
could bring great surprises in a not so far future.

\section*{Acknowledgements}
The authors would like to thank Raúl Carballo-Rubio and Luis J. Garay for enlightening discussions. 
Financial support was provided by the Spanish Government through the projects PID2020-118159GB-C43, PID2020-118159GB-C44 (with FEDER contribution), and by the Junta de Andaluc\'{\i}a through the project FQM219. 
The authors acknowledge financial support from the grant CEX2021-001131-S funded by MCIN/AEI/ 10.13039/501100011033.
%============================================================

%============================================================

%============================================================

\newpage
\begin{thebibliography}{99}
%============================================================

%\cite{EventHorizonTelescope:2022xqj}
\bibitem{EventHorizonTelescope2022}
K.~Akiyama \textit{et al.}, 
[Event Horizon Telescope],
%``First Sagittarius A* Event Horizon Telescope Results. VI. Testing the Black Hole Metric,''
Astrophys. J. Lett. \textbf{930}, no.2, L17 (2022).
%doi:10.3847/2041-8213/ac6756
%164 citations counted in INSPIRE as of 28 Mar 2023



%\cite{Carballoatal2018c}
\bibitem{Carballoetal2018c}
R.~Carballo-Rubio, P.~Kumar and W.~Lu,
%``Seeking observational evidence for the formation of trapping horizons in astrophysical black holes,''
Phys. Rev. D \textbf{97}, 123012 (2018).
%doi:10.1103/PhysRevD.97.123012
%[arXiv:1804.00663 [gr-qc]].
%19 citations counted in INSPIRE as of 28 Mar 2023


%\cite{Carballo-Rubio:2022aed}
\bibitem{Carballo-Rubioetal2022}
R.~Carballo-Rubio, V.~Cardoso and Z.~Younsi,
%``Toward very large baseline interferometry observations of black hole structure,''
Phys. Rev. D \textbf{106}, 084038 (2022).
%doi:10.1103/PhysRevD.106.084038
%[arXiv:2208.00704 [gr-qc]].
%8 citations counted in INSPIRE as of 28 Mar 2023

%\cite{Konoplya:2016pmh}
%\bibitem{KonoplyaZhidenko2016}
%R.~Konoplya and A.~Zhidenko,
%``Detection of gravitational waves from black holes: Is there a window for alternative theories?,''
%Phys. Lett. B \textbf{756}, 350-353 (2016)
%doi:10.1016/j.physletb.2016.03.044
%[arXiv:1602.04738 [gr-qc]].
%161 citations counted in INSPIRE as of 28 Mar 2023

%\cite{Abramowicz:2002vt}
%\bibitem{Abramowiczetal2002}
%M.~A.~Abramowicz, W.~Kluzniak and J.~P.~Lasota,
%``No observational proof of the black hole event-horizon,''
%Astron. Astrophys. \textbf{396}, L31-L34 (2002)
%doi:10.1051/0004-6361:20021645
%[arXiv:astro-ph/0207270 [astro-ph]].
%189 citations counted in INSPIRE as of 28 Mar 2023

%\cite{Oppenheimer:1939ne}
\bibitem{OppenheimerVolkoff1939}
J.~R.~Oppenheimer and G.~M.~Volkoff,
%``On massive neutron cores,''
Phys. Rev. \textbf{55}, 374-381 (1939).
%doi:10.1103/PhysRev.55.374
%2430 citations counted in INSPIRE as of 28 Mar 2023

%\cite{Buchdahl:1959zz}
\bibitem{Buchdahl1959}
H.~A.~Buchdahl,
%``General Relativistic Fluid Spheres,''
Phys. Rev. \textbf{116}, 1027 (1959).
%doi:10.1103/PhysRev.116.1027
%811 citations counted in INSPIRE as of 21 Mar 2023

%\cite{Howard:1984qp}
\bibitem{HowardCandelas1984}
K.~W.~Howard and P.~Candelas,
%``QUANTUM STRESS TENSOR IN SCHWARZSCHILD SPACE-TIME,''
Phys. Rev. Lett. \textbf{53}, 403-406 (1984).
%doi:10.1103/PhysRevLett.53.403
%121 citations counted in INSPIRE as of 21 Mar 2023

%\cite{Anderson:1994hg}
\bibitem{Andersonetal1994}
P.~R.~Anderson, W.~A.~Hiscock and D.~A.~Samuel,
%``Stress - energy tensor of quantized scalar fields in static spherically symmetric space-times,''
Phys. Rev. D \textbf{51}, 4337-4358 (1995).
%doi:10.1103/PhysRevD.51.4337
%199 citations counted in INSPIRE as of 21 Mar 2023

%\cite{Levi:2016exv}
\bibitem{Levietal2016}
A.~Levi, E.~Eilon, A.~Ori and M.~van de Meent,
%``Renormalized stress-energy tensor of an evaporating spinning black hole,''
Phys. Rev. Lett. \textbf{118}, 141102 (2017).
%doi:10.1103/PhysRevLett.118.141102
%[arXiv:1610.04848 [gr-qc]].
%32 citations counted in INSPIRE as of 21 Mar 2023

%\cite{Simon:1990jn}
\bibitem{Simon1990}
J.~Z.~Simon,
%``The Stability of flat space, semiclassical gravity, and higher derivatives,''
Phys. Rev. D \textbf{43}, 3308-3316 (1991).
%doi:10.1103/PhysRevD.43.3308
%108 citations counted in INSPIRE as of 21 Mar 2023

%\cite{Flanagan:1996gw}
\bibitem{FlanaganWald1996}
E.~E.~Flanagan and R.~M.~Wald,
%``Does back reaction enforce the averaged null energy condition in semiclassical gravity?,''
Phys. Rev. D \textbf{54}, 6233-6283 (1996).
%doi:10.1103/PhysRevD.54.6233
%[arXiv:gr-qc/9602052 [gr-qc]].
%258 citations counted in INSPIRE as of 21 Mar 2023



%\cite{Polyakov:1981rd}
\bibitem{Polyakov1981}
A.~M.~Polyakov,
%``Quantum Geometry of Bosonic Strings,''
Phys. Lett. B \textbf{103}, 207-210 (1981).
%doi:10.1016/0370-2693(81)90743-7
%2992 citations counted in INSPIRE as of 21 Mar 2023

%\cite{Arrechea:2021pvg}
\bibitem{Arrecheaetal2021}
J.~Arrechea, C.~Barcel\'o, R.~Carballo-Rubio and L.~J.~Garay,
%``Semiclassical constant-density spheres in a regularized Polyakov approximation,''
Phys. Rev. D \textbf{104}, 084071 (2021).
%doi:10.1103/PhysRevD.104.084071
%[arXiv:2105.11261 [gr-qc]].
%11 citations counted in INSPIRE as of 21 Mar 2023


\bibitem{Thesis}
J.~Arrechea, \textit{PhD Thesis}, University of Granada (2023) (available upon request).\\
J.~Arrechea, C.~Barceló, R.~Carballo-Rubio and L.~J.~Garay, \textit{forthcoming publication}.

%\cite{Arrechea:2019jgx}
\bibitem{Arrecheaetal2019}
J.~Arrechea, C.~Barcel\'o, R.~Carballo-Rubio and L.~J.~Garay,
%``Schwarzschild geometry counterpart in semiclassical gravity,''
Phys. Rev. D \textbf{101}, 064059 (2020).
%doi:10.1103/PhysRevD.101.064059
%[arXiv:1911.03213 [gr-qc]].
%9 citations counted in INSPIRE as of 21 Mar 2023

%\cite{Arrechea:2022dvy}
\bibitem{Arrecheaetal2022b}
J.~Arrechea, C.~Barcel\'o, R.~Carballo-Rubio and L.~J.~Garay,
%``Asymptotically flat vacuum solutions in order-reduced semiclassical gravity,''
arXiv:2212.09375.
%0 citations counted in INSPIRE as of 21 Mar 2023


%\cite{Misner:1964je}
\bibitem{MisnerSharp1964}
C.~W.~Misner and D.~H.~Sharp,
%``Relativistic equations for adiabatic, spherically symmetric gravitational collapse,''
Phys. Rev. \textbf{136}, B571-B576 (1964).
%doi:10.1103/PhysRev.136.B571
%957 citations counted in INSPIRE as of 21 Mar 2023

%\cite{Hernandez:1966zia}
\bibitem{HernandezMisner1966}
W.~C.~Hernandez and C.~W.~Misner,
%``Observer Time as a Coordinate in Relativistic Spherical Hydrodynamics,''
Astrophys. J. \textbf{143}, 452 (1966).
%doi:10.1086/148525
%163 citations counted in INSPIRE as of 21 Mar 2023

%\cite{Schwarzschild:1916ae}
\bibitem{Schwarzschild1916}
K.~Schwarzschild,
%``On the gravitational field of a sphere of incompressible fluid according to Einstein's theory,''
Sitzungsber. Preuss. Akad. Wiss. Berlin (Math. Phys.) \textbf{1916}, 424-434 (1916).
%134 citations counted in INSPIRE as of 29 Mar 2023

%\cite{Urbano:2018nrs}
\bibitem{UrbanoVeermae2018}
A.~Urbano and H.~Veerm\"ae,
%``On gravitational echoes from ultracompact exotic stars,''
JCAP \textbf{04}, 011 (2019).
%doi:10.1088/1475-7516/2019/04/011
%[arXiv:1810.07137 [gr-qc]].
%45 citations counted in INSPIRE as of 21 Mar 2023

\bibitem{Hiscock1988}
W.~A.~Hiscock,
Phys. Rev. D \textbf{37}, 2142-2150 (1988).
%doi: 10.1103/PhysRevD.37.2142

%\cite{Reyes:2023fde}
\bibitem{ReyesTomaselli2023}
I.~A.~Reyes and G.~M.~Tomaselli,
%``Compact stars in Quantum Field Theory,''
arXiv:2301.00826.
%0 citations counted in INSPIRE as of 21 Mar 2023



\bibitem{DaviesFulling1977}
P.~C.~W.~Davies, S.~A.~Fulling,
Proc. R. Soc. Lond. A \textbf{354} 59–77 (1977).%doi:10.1098/rspa.1977.0056


%\cite{Parentani:1994ij}
\bibitem{ParentaniPiran1994}
R.~Parentani and T.~Piran,
%``The Internal geometry of an evaporating black hole,''
Phys. Rev. Lett. \textbf{73}, 2805-2808 (1994).
%doi:10.1103/PhysRevLett.73.2805
%[arXiv:hep-th/9405007 [hep-th]].
%89 citations counted in INSPIRE as of 21 Mar 2023
%\cite{Fabbri:2005zn}
\bibitem{Fabbrietal2005}
A.~Fabbri, S.~Farese, J.~Navarro-Salas, G.~J.~Olmo and H.~Sanchis-Alepuz,
%``Semiclassical zero-temperature corrections to Schwarzschild spacetime and holography,''
Phys. Rev. D \textbf{73}, 104023 (2006).
%doi:10.1103/PhysRevD.73.104023
%[arXiv:hep-th/0512167 [hep-th]].
%28 citations counted in INSPIRE as of 21 Mar 2023

%\cite{Chakraborty:2015nwa}
\bibitem{Chakrabortyetal2015}
S.~Chakraborty, S.~Singh and T.~Padmanabhan,
%``A quantum peek inside the black hole event horizon,''
JHEP \textbf{06}, 192 (2015).
%doi:10.1007/JHEP06(2015)192
%[arXiv:1503.01774 [gr-qc]].
%21 citations counted in INSPIRE as of 21 Mar 2023
%\cite{Arrechea:2021xkp}


\bibitem{Arrecheaetal2022}
J.~Arrechea, C.~Barcel\'o, R.~Carballo-Rubio and L.~J.~Garay,
%``Semiclassical relativistic stars,''
Sci. Rep. \textbf{12}, 15958 (2022)
%doi:10.1038/s41598-022-19836-8
%[arXiv:2110.15808 [gr-qc]].
%7 citations counted in INSPIRE as of 21 Mar 2023


%\cite{Landau:1975pou}
\bibitem{LandauLifschits1975}
L.~D.~Landau and E.~M.~Lifschits,
{\textit{The Classical Theory of Fields}},
Pergamon Press (1975).
%ISBN 978-0-08-018176-9
%88 citations counted in INSPIRE as of 27 Mar 2023

%\cite{Parker:1993dk}
\bibitem{ParkerSimon1993}
L.~Parker and J.~Z.~Simon,
%``Einstein equation with quantum corrections reduced to second order,''
Phys. Rev. D \textbf{47}, 1339-1355 (1993).
%doi:10.1103/PhysRevD.47.1339
%[arXiv:gr-qc/9211002 [gr-qc]].
%134 citations counted in INSPIRE as of 27 Mar 2023


%\cite{EventHorizonTelescope:2019dse}
\bibitem{EventHorizonTelescope2019}
K.~Akiyama \textit{et al.} [Event Horizon Telescope],
%``First M87 Event Horizon Telescope Results. I. The Shadow of the Supermassive Black Hole,''
Astrophys. J. Lett. \textbf{875}, L1 (2019).
%doi:10.3847/2041-8213/ab0ec7
%[arXiv:1906.11238 [astro-ph.GA]].
%2130 citations counted in INSPIRE as of 15 Mar 2023

\bibitem{LIGOScientific2016}
B.~P.~Abbott \textit{et al.} [LIGO Scientific and Virgo],
%``Observation of Gravitational Waves from a Binary Black Hole Merger,''
Phys. Rev. Lett. \textbf{116}, 061102 (2016).
%doi:10.1103/PhysRevLett.116.061102
%[arXiv:1602.03837 [gr-qc]].
%9133 citations counted in INSPIRE as of 15 Mar 2023



%\cite{Abedi:2016hgu}
\bibitem{Abedietal2016}
J.~Abedi, H.~Dykaar and N.~Afshordi,
%``Echoes from the Abyss: Tentative evidence for Planck-scale structure at black hole horizons,''
Phys. Rev. D \textbf{96}, 082004 (2017).
%doi:10.1103/PhysRevD.96.082004
%[arXiv:1612.00266 [gr-qc]].
%285 citations counted in INSPIRE as of 27 Mar 2023

%============================================================
\end{thebibliography}
\end{document}